# Learning and Upgrading in Global Value Chains: An Analysis of India's Manufacturing Sector

Sourish Dutta

PhD Student, 2016-20

Centre for Development Studies

Trivandrum, Kerala

Proposal for Thesis Research in Partial Fulfillment of the Requirements for the Degree of Doctor of Philosophy in Economics from the Jawaharlal Nehru University, New Delhi

\_\_\_\_\_\_\_\_\_\_\_\_\_\_\_\_\_\_\_\_\_\_\_\_\_             \_\_\_\_\_\_\_\_\_\_\_\_\_\_\_\_\_\_\_\_\_\_\_\_\_\_

Prof. K. J. Joseph                                          Dr. Parameswaran. M

\_\_\_\_\_\_\_\_\_\_\_\_\_\_\_\_\_\_\_\_\_\_\_\_

Sourish Dutta

# 1 BACKGROUND

Globalisation does not merely mean a quantitative expansion in international economic activity; it also indicates a qualitative shift (Milberg & Winkler 2013). Deepening of trade liberalisation with continuing reduction of transportation cost, revolution in the information and communications technology (ICT), and some recent political developments are expanding the reach of globalisation through a gradual fragmentation and distribution of production processes across countries (Antràs 2016). Nowadays most of the products are composed of different designs and components produced by many suppliers (firms) spread across various countries with its profound impact similar to what Smith (1776) observed regarding the division of labour.

The global production structure, a system of creating values in geographically separated tasks or phases, is eventually forming an extensive network of economic values or value added that explains the changing nature of international trade-growth-development links. It is termed as the global value chains (GVCs), creating the nexus among firms, workers and consumers around the globe. In general, from the industrial organisation perspective, value chains describe the sequence of productive (value-added) activities that capital and labour (or firms and workers) perform to bring a good or service from its conception to end use and beyond (Porter 1985, Sturgeon 2001). "Value chain analysis" aims at identifying bottlenecks and opportunities between different stages of production and tasks. It includes activities such as design, administrative services, manufacturing, assembling, marketing, distribution and customer services. In the context of globalisation, these tasks constitute a value chain as they have been carried out in inter-firm and intra-firm networks on a global scale (Gereffi et al. 2001, 2005). These value chains can be regional if the scope of production takes place within the same geographic region.

The phenomenon of global value chains (GVCs) indicates a division of labour type production structure in which tasks and business functions are distributed among several companies, globally, or regionally (Grossman & Rossi-Hansberg 2008). The critical features of GVCs are therefore the international dimension of the production process and the "contractualisation" of buyer and seller relationships, often across international borders (Antràs 2016). As a result, these international production networks are highly complex regarding geography, technology, and the different types of firms involved (from large retailers and highly large-scale mechanised manufacturers to small home-based production). Sometimes it may be impossible even to identify all the countries that are involved or the extent of their involvement (Gereffi & Fernandez-Stark 2016). However, the recent development of OECD-WTO's Trade in Value Added (TiVA) data represents a fundamental step forward in understanding GVC trade. Grossman & Rossi-Hansberg (2008, 2012) rightly point out that the different tasks, rather than sectors, define the specialisation of countries in the value chains.

In this global bandwagon, India has not been left behind. India's participation in GVCs in the last two decades concerning the foreign content of its exports was more than double from less than 10% in 1995 to 24% in 2011. It has been argued that the increased participation GVCs be associated with the hollowing out of Indian manufacturing. Indian industrial sector is experiencing a rising output but falling value-added in total production (declining real value added growth and employment elasticity) with the trend becoming more pronounced since the mid-1990s. Besides this, dualism concerning 'formal', 'informal', and 'missing-middle' along with limiting regulations pose unique challenges to the growth of India's manufacturing (Banga 2014*b*). As Indian industries are facing an intense competition (domestic as well as external) linked with the global production sharing, the obvious increased use of imported inputs has caused a generalised decline in national value-added share for merchandise and total exports (Banga 2014*a*, Goldar et al. 2017, Veeramani & Dhir 2017). Although India's output grows and exports rise, if the domestic value added does not rise, then there would be no noticeable production-linked gains like employment generation, technology upgrading, and skill development (Banga 2014*b*). It requires much more value added from India's potential productive factors and upgrading quality & quantity of those factors with a proper distributional aspect of socioeconomic opportunities and outcomes (Banga 2014*a*).

Indeed, Milberg & Winkler (2013) rightly suggest a shift in emphasis from static efficiency gains (resulting from specialisation and exchange) to the questions of the sources and uses of profits for firm investment, employment demand, and innovation to understand the welfare implications of an economy's global production networks (dynamic gains). In this context, Stiglitz & Greenwald (2014) have asked a central question about determinants of a country's dynamic comparative advantage... "*It has become conventional wisdom to emphasize that what matters is not static comparative advantage but dynamic comparative advantage. Korea did not have a comparative advantage in producing semiconductors when it embarked on its transition. Its static comparative advantage was in the production of rice. Had it followed its static comparative advantage (as many* *neoclassical economists had recommended), then that might still be its comparative advantage; it might be the best rice grower in the world, but it would still be poor. But a country's dynamic comparative advantage is endogenous, a result of what it does. There seems to be a circularity here. The central question is, what should a country do today to create its dynamic comparative advantage?"* According to their perspective, the most crucial endowment for this redefined comparative advantage is an economy's state of knowledge and learning capabilities. In simple words, upgrading in GVCs is a multidimensional process that seeks to upgrade the economic conditions (profits, employment, skills) and social conditions (working conditions, low incomes, education system) of a firm, industry or group of workers Keane (2017). Milberg and Winkler (2013) describe upgrading within GVCs as being synonymous with economic development. From this perspective, upgrading involves a learning process through which firms acquire knowledge and skills – often through their relationships with other enterprises in the value chain or through supporting markets – that can be translated into innovations or improvements that increase the value of their goods or services Keane (2017). This process requires a country's policies to be shaped to take benefit of its comparative advantage in knowledge accumulation and learning capabilities, including its ability to learn and to learn to learn, concerning its competitors, and to help develop those capabilities further (Stiglitz and Greenwald 2014). Now it is pertinent to ask how an economy's production processes, producers and employees are connecting to the world trade and capturing the dynamic gains out of it? It implies that what is an economy's position in GVCs? What can an economy do now to ensure that it increases its share of domestic value added in these GVCs? How can an economy leverage its participation in GVCs to increase innovation, productivity and inclusivity? In summary, the core issue is (Lundvall 2016): Under what circumstances will the participation in GVCs contribute to learning and upgrading at the level of the firm, at the level of a sector and to economic and social development at the national level? This research would analyse this issue through three phases of knowledge for GVC integration (Taglioni & Winkler 2016), i.e. knowledge of entering GVCs, knowledge of expanding and strengthening GVC participation, and knowledge of turning GVC participation into sustainable development.

## 2 REVIEW OF THE LITERATURE

This section tries to understand this new phenomenon by exploring the development of relevant studies. It surveys mainly those strands of research which explicitly examine the vertical (supply–use) relationships of global production sharing and their impact on the distribution of total value added among the economic agents, which is to be the heart of GVC studies. A particular concern is a difficulty of creating a boundary between GVC studies and international trade literature, as these two areas overlap in many respects and the relevant work is often cross-cited.

### 2.1 Theoretical Literature

The two-century-old Ricardian foundation of international trade (i.e. comparative advantage from Heckscher-Ohlin to Samuelson) has based on three standard assumptions, viz. producers operate at constant returns to scale in perfectly competitive markets, an industry consists of homogeneous manufacturers, and countries trade only final products. This traditional foundation was improved with a new trade theory by Krugman (1979, 1980), Helpman & Krugman (1985). The new trade theory considers increasing returns to scale in production technology (paired with the love of variety) under imperfectly competitive markets. It is an explanation of the intra-industry trade between countries with similar technology and resource endowment, which cannot be explained by the conventional notion of comparative advantage. This theoretical breakthrough paved several development pathways in the days that followed. We know that the empirical findings on intra-industry trade by Grubel & Lloyd (1975) had helped to give birth of the New Trade Theory to improve the first assumption of the old trade theory. Similarly, the second conventional assumption of homogeneous producers was reconsidered (detailed examination of firm-level data) by Bernard et al. (1995), Bernard & Jensen (1999) to revealed substantial heterogeneity in firm productivity between exporters and non-exporters in a given industry. It had helped to generate the New-New Trade Theory of Melitz (2003). This theory considers the mechanism of a firm's endogenous selection on market entry or exit assuming a fixed cost of entering export activities. As a result, industry became an inappropriate economic unit for the study of international trade. Echoing Inomata (2017), a third wave of reconstructing classical theory is now underway, and the literature on GVCs is gradually linking to this development strand. As we have already discussed in the first section, the revolutionary advancement of transportation and information and communication technology (ICT) have made production processes to be "sliced" into several production segments, each corresponding to a particular task, viz. design, parts procurement, assembly, and distribution. These segments are relocated across national borders to the places where the tasks can be executed most economically. Thus the core of the literature today is not only the movement of final goods and services, but also the cross-border transfer of tasks or the value added generated by these tasks.

Jones & Kierzkowski (2001) introduced a theoretical framework of the fragmented productive activities through a simple model of outsourcing and find out the factors that affect its degree & form. It implies that (Other things being equal) the production process will be more prone to international fragmentation if the following three conditions are there. Firstly, the targeted market is large enough to absorb the increased supply of goods from the organisation of more efficient divisions of labour across countries. Secondly, the costs of linking the production segments in different countries are less inhibitive. Thirdly, the countries in the production network are more heterogeneous regarding their factor costs, so there is plenty of choice for offshoring firms to exploit comparative advantage. This model was further developed to address income distribution and welfare leading to the industrial hollowing-out problem, i.e. moving some tasks to foreign countries frees up the domestic labour that would otherwise carry out these tasks. So it affects by increasing labour supply in the market lowering the real wages of offshored labour or losing domestic jobs when wages are sticky. Traditionally, the effect of international trade on the labour market has been thought regarding a resource movement between industrial sectors caused by import competition, without much attention to the change in the within-sector composition of different types of labour. Newer globalisation literature focuses on this point, recognising that offshoring is a cross-border movement of a production activity corresponding to a task to a particular type and skill of labour. It had also influenced the following range of studies by increasing observations of trade in intermediate goods (Feenstra & Hanson 1996, Campa & Goldberg 1997, Yeats 1998). These studies brought about the further elaboration of critical concepts such as first and second unbundling of production activities (Baldwin 2006), and trade in tasks with an improved coordination capability between the firm's headquarters & its foreign suppliers through transportation and communication technologies (Grossman & Rossi-Hansberg 2008).

It leads to a productivity effect equivalent to the consequence of factor-augmenting technological progress, so it can bring a positive impact on the employment of domestic workers (across all industries) whose task levels are similar to those of offshored labour. However, Some tasks (such as those akin to the codified description) are easy to offshore, while others (such as those relying on tacit personal knowledge) are not (Blinder et al. 2009, Grossman & Rossi-Hansberg 2012). This trade in tasks from developed countries to developing countries are associated with transfers of some phases of production processes that are considered more skill-intensive by the level of developing countries but less skill-intensive by the level for developed countries. Accordingly, the demand for labour becomes skewed toward high-skilled workers in the light of the respective skill level of each economy, so the relative wages of low-skilled labour fall in both developed and developing countries. However, the declining relative wage does not primarily make unskilled workers worse off because from a general equilibrium perspective, the increased supply of goods to the market brought about by the more delicate division of labour may lower the goods prices of both countries through trade, perhaps offsetting the nominal wage reduction.

**2.2 Empirical Literature**

The recent availability of relevant data and statistics, especially inter-country input-output tables and firm-level data, has led to the rapid progress of empirical literature on GVCs. Earlier value-added analyses were based on firms' business records (Dedrick et al. 2008, Xing & Detert 2010), and the essence of these studies was to investigate how firms benefit from technological innovation through global production sharing. However, it came to elucidate a separate and even more alarming question about the validity of conventional trade statistics based on gross values. Although the product-level studies are beneficial for drawing the actual structure of production chains, they do not provide a macro picture. However, now those are complemented by input-output analysis, in which various GVC metrics were devised using inter-country input-output databases, viz. trade in value added (Johnson & Noguera 2012, 2016) and supply chain length (Dietzenbacher et al. 2005, Fally 2011). The inter-country input-output table provides a comprehensive map of international transactions of goods and services in a massive data-set that combines the national input-output tables of various countries at a given point of time. Because the tables contain information on supply–use relations between industries and across countries— Which are absent from foreign trade statistics— it is possible to identify the vertical structure of international production sharing (Hummels et al. 2001). Moreover, unlike the product-level approach, the input-output analysis covers an entire set of industries that make up an economic system, thus enabling the measurement of cross-border value flows for a country or region. Theoretically, such analysis can track the value-added generation process of every product in every country at every production stage (Koopman et al. 2012, 2014, Timmer et al. 2014). Antras & Helpman (2004) integrated both the New Trade Theory (increasing returns to scale) and the New-New Trade Theory (firm heterogeneity) in the context of contract theory. They examined the impact of within-sector heterogeneity in firm productivity on its globalisation decision. The model suggests that different degrees of entry cost to global activities bring about the productivity ranking among firms on the choice of globalisation modes. The most productive firms would choose to undertake FDI, the next most productive firms would take part in arm's length offshoring, and so on down to the least productive firms, which would participate only in domestic procurement. Antràs & Chor (2013) carried over this model to further incorporating the methodological progress in input-output economics by considering a technological ordering of production stages, a crucial attribute of value chains, to address the general make-or-buy question for each segment of a production process along a value chain. However, the input-output approach is not free from weaknesses. The usual input-output table is based on industrial categories, and hence the value-added of a specific production activity such as product design or assembly cannot be identified (Sturgeon et al. 2013). In India, a few studies provide the time series estimates of the domestic value-added content of India's merchandise and services exports. Veeramani & Dhir (2017) make use of the official input-output tables for the benchmark years 1998-99, 2003-04, 2007-08 as well as the Supply Use Tables for the years 2011-12 and 2012-13. However, Goldar et al. (2017) and Banga (2013) have studied the increased use of imported inputs causing a generalised decline in domestic value-added share for merchandise and total exports (using OECD-WTO TiVA and WIOD databases) in India along with other critical emerging nations.

Internationally fragmented production is not new. For decades, low- and middle-income countries (LMICs) have imported parts from countries with more advanced technology, although only for the assembly of domestically sold goods. As the goods produced were not part of a global production network, flows of know-how were less intense. The characteristic of GVCs from a development perspective is that factories in LMICs have become full-fledged participants in international production networks. They are no longer just importing final parts for assembly in domestic sales. They are exporting goods, parts, components, and services customised to the needs of the intended buyers and used in some of the most sophisticated products on the planet (Taglioni & Winkler 2016).

Nowadays the GVC-enabled flow of know-how from high-income countries to LMICs is determining the industrial development. LMICs can now industrialise by joining GVCs without the need to build their value chain from inception, as Japan and the Republic of Korea had to do in the twentieth century (Baldwin et al. 2012). That enables LMICs to focus on specific production activities in the value chain rather than producing the whole product, thereby lowering the threshold and costs for industrial development. LMICs can benefit from the foreign-originated intellectual property; trademarks; operational, managerial, and business practices; marketing expertise; and organisational models. Countries have to understand the opportunities that GVCs offer and adopt the appropriate policies to mitigate the risks associated with them have the possibility—through GVCs—to boost employment and productivity in all their agriculture, manufacturing, and services production. Job creation and labour productivity growth are sometimes viewed as competing goals, as higher labour productivity enables firms to produce a more significant amount of value added without necessarily increasing the number of workers at the same rate (static productivity effects). Research shows that GVC integration leads to higher net jobs, but lower employment intensity (Calì et al. 2016, 2015). It has a high potential for productivity gains via several transmission channels (dynamic productivity effects), as discussed later, which goes in hand with increased labour demand caused by more vertical specialisation and higher output in GVCs.

## 3 DEFINITION, RATIONALE AND SCOPE OF THE STUDY

Understanding the development process through competitive GVC participation needs a proper examination of the main factors behind the GVC integration and its measurements. Assessment of India's GVC participation requires three factors:

1. Functions in GVCs: considering India's performance as a buyer as well as a seller in the global market.
2. Specialisation and domestic value-added contribution: trends of specialisation in low or high value-added activities, and patterns of upgrading and development through GVCs.

3. Position in GVC network and type of GVC node: hub, incoming spoke, or outgoing spoke; clustering properties; and centrality in the global network

The multidimensionality of GVCs can be grasped by looking at the relationships between flows of goods, services and flows of factors of production (workers, ideas, and investments). Hence, we have to go beyond value added to look at the actors in GVCs and assess the effects of GVCs on jobs and wages (Taglioni & Winkler 2016).

**3.1 Functions in GVCs: Buyer's and Seller's Perspectives**

Traditional trade theories assume that the whole production process of a product is taken place in one country and marketed in another. However, the notion of GVC trade is different – quantifying how much of India's export value is contributed by foreign countries and how much India is adding value to exports and final demand in third countries. The term GVC trade typically refers to value-added trade produced goods and related services, but more generally it also includes imported raw components used in exports. The basic concept is "import to export" or I2E (Baldwin & Lopez-Gonzalez 2015). For example, On the sourcing side (buyer's perspective), it indicates that India is buying parts from Japan in a GVC. On the sales side (seller's perspective), it indicates that India is using those parts in its exports to that GVC. Import to export on the sales and sourcing sides is linked to the bilateral concepts of backward and forward vertical specialisation (González & Holmes 2011), in which "backward" refers to sourcing and "forward" to sales. Actualisation of the I2E concept can be made a distinction between the seller's and buyer's sides of GVC participation. In many cases, countries are GVC buyers and GVC sellers, but that distinction reflects the difference in economic mechanisms and determinants that lead to a country's satisfactory performance in absorbing valuable foreign value added compared with growing domestic value embodied in GVC trade flows. Taglioni & Winkler (2016) consider three types of buyer roles in GVCs: input purchases (1) for the production of final exports, (2) for the production of intermediate inputs in the value chain, and (3) for assembly. There are also three main seller functions: supply of (1) turnkey components, (2) primary inputs, and (3) other inputs.

The flows of various goods, services, people, ideas, and capital, which are predominantly associated with the buyer's or seller's role, are more easily discussed by first focusing on the buyer's or seller's functions separately and then considering them jointly. That evaluation is more easily actionable from the policy angle. If for example, the domestic value chain is found to be short, or little transformation is taking place domestically, the supply-side bottlenecks and opportunities for expansion on the buying side could be more readily identified than those on the selling side (Taglioni & Winkler 2016).

### 3.2 Specialisation and Value Addition

It is evident that the most important things for the economics of GVCs are a generation of value addition and its growth over the time. Although value addition is a function of productivity, it is also linked to the quantity, diversity, and quality of tasks and activities in which an economy specialises depending upon the level of innovation capacity. The variety of activities in a value chain is extensive. The activities range from manufacturing inputs, outputs, and assembly operations to logistics, marketing, sales, and a range of other service activities. Moreover, there are activities as diverse as the production of other inputs, machinery, and equipment, as well as R&D, technological development, and functions aimed at organising the firm's infrastructure, human resource management, and procurement. Broadly, the value-added content of such activities and tasks tends to grow as the level of technological and know-how skills to perform the task increase.

In a world dominated by complex and fragmented production processes, economic development can occur through industrial or economic upgrading and densification. Economic upgrading, often referred to as industrial upgrading or merely upgrading, is defined as the ability of producers to make better or new products, to make products more efficiently, or to move into more skilled activities (Pietrobelli & Rabellotti 2006). In other words, this upgrading means increasingly embracing higher value-added production with the contribution of better skills and know-how, capital and technology, and processes. It can be realised in the form of (1) process upgrading through improving the organisational process, (2) product upgrading through introducing new products or improving existing products, (3) functional upgrading through changing the mix of activities, and (4) inter-sectoral upgrading through moving to a new value chain (Cattaneo & Miroudot 2013, Humphrey 2004, Humphrey & Schmitz 2002, Kaplinsky & Morris 2001). Densification involves engaging more local actors (firms and workers) in the GVC network reinforcing living standards, including employment, wages, working conditions, economic rights, gender equality, economic security, and protection of the environment. In some cases, this could mean that performing lower value-added activities on a larger scale can generate significant value addition for India. Therefore, engineering equitable distribution of opportunities and outcome aspects are essential to the analysis of the extent to which industrial or economic upgrading is associated with social upgrading (Milberg & Winkler 2013). There are several variables formulated to measure economic and social upgrading at different levels of analysis: the nation, the sector or GVC, and the firm or the plant. Those are:

| Level of aggregation | Economic Upgrading | Social Upgrading |
|---|---|---|
| Country | Productivity growth<br>Value-added growth<br>Profits growth<br>Export growth<br>Growth in export market share<br>Unit value growth of output<br>Unit value growth of exports<br>Reduced relative unit labour costs | Wage growth<br>Employment growth<br>Growth in labour share<br>Formal employment<br>Youth employment<br>Gender equality<br>Poverty reduction<br>Share of wage employment<br>Improved labour standards<br>Regulation of monitoring<br>Improved political rights<br>Human development index |
| Sector or GVC | Productivity growth<br>Value-added growth<br>Profits growth<br>Export growth<br>Growth in export market share<br>Unit value growth of output<br>Unit value growth of exports<br>Reduced relative unit labour costs<br>Increased capital intensity<br>Increased skill intensity of functions<br>Increased skill intensity of employment<br>Increased skill intensity of exports | Wage growth<br>Employment growth<br>Labour standards |
| Firm | Skill intensity of functions<br>Skills to manage the supply chain<br>Composition of jobs<br>Capital intensity/mechanisation<br>Product, process, functional, chain | Standards in plant monitoring<br>Number of workers per job |

## 3.3 Position in GVC Networks and Type of GVC Node

In the context of complexity and multidimensionality of GVCs, network analysis can trace the overall performance of India's different actors and trade links (using domestic value added in gross exports). This assessment can be executed by creating a trade network topology, consisting of a set of centrality measures that capture various aspects of the network (Santoni & Taglioni 2015, Amador & di Mauro 2015). The appropriate steps are:

- Strength: Sum of values of inflows or outflows. The use of normalised link weights implies that the values for in-strength and out-strength report the market share of India. The values show that usual market shares are a particular case of network centrality measures when considering only first-order connectedness.

- Closeness: A measure of how close (topological distance) a node is to all other nodes. In general terms, the concept of distance in network analysis is related to the number of steps needed for some node to "reach" another network node. In the case of weighted networks, not the number of steps but the value of the links (the inverse of link value) is considered; the most substantial flows result from a shorter distance.

- Centrality: Expresses the idea that the influence of India (as a node) is proportional to the influence of its neighbours (or peers). It is the most representative measure of the network and captures the links and their closeness or proximity. The centrality can be computed from the buyer's or seller's perspective (Bonacich 1987).

- Clustering: Expresses the transitivity of the network, measuring how much neighbours (or peers) of India are connected to each other. It captures whether India is strong because it trades a lot with other countries that are also strong.

These metrics can also illustrate other types of flows (for example, parts, components, services, or FDI) or flows in individual sectors or of specific products (Taglioni & Winkler 2016). In fact, in GVCs, input-output links can generate a cascade effect induced by the propagation of micro-shocks through the production network (Acemoglu et al. 2012, 2015, 2016).

Echoing Antràs & Rossi-Hansberg (2009), there is a much scope to study the dynamic impact of the international organisation of production on the evolution of knowledge, the distribution of skills, and other country-specific characteristics. This research scope indeed brings my attention to the synergies between different areas of GVCs. Now it would be logical to examine India's efforts to identify trade competitiveness (mainly measured in value added), performance in GVC integration (economic & social upgrading), and the role of country characteristics, including the business climate, investment climate, and drivers of competitiveness across economic, regulatory, operational, and infrastructural dimensions. This study is all about the global value chains of India's manufacturing industries. Of course, these do not only contain activities in the manufacturing sector, but also in other sectors, i.e. agriculture, utilities, and business services that provide inputs at any stage of the production process of manufactures. These indirect sectoral contributions are sizeable and will be explicitly computed through the modelling of input-output linkages across industries. This study could be developed through a most extensive range of available and applicable methodologies followed by an in-depth assessment and discussion of the identified challenges and opportunities.

## 4 RESEARCH QUESTIONS & HYPOTHESES

To analyse India's position, functions, specialisation & value addition of manufacturing GVCs, it is required to quantify the extent, drivers, and impacts of India's Manufacturing links in GVCs. This overall broad objective can be transformed into three fundamental questions:

1. **What is the Extent of India's Manufacturing Links in GVCs?**

   - This query proposes to identify potential manufacturing industries in GVCs through analysing the informed classifications (by final use and chain category), backward integration, forward integration, overall GVC participation, GVC position, and GVC length indices, network metrics & visualisation, service integration, and firm-level measures to capture the critical features of firm heterogeneity in GVC participation.

2. **What are the Determinants of India's Manufacturing Links in GVCs?**

   - Regarding the determinants of India's manufacturing GVCs, the query would begin with the decomposition of gross export growth, which is the first analysis of where the growth of the value added embodied in gross exports generated regarding the country of origin (i.e. foreign versus domestic) and sector ( i.e. intra-sector versus inter-sector).

- This query would then look into more specific determinants of GVC engagement at the industry and firm levels based on the knowledge of entering GVCs. Industry level determinants are the presence of export processing zone, openness to world trade and FDI, connectivity to global markets, competitiveness regarding unit labour costs and labour productivity, drivers of investment, and the quality of domestic value chains and the services infrastructure. Firm-level determinants include size, age, foreign ownership status, workers' skills, productivity, and institutional variables (logistics performance, competitiveness, and absorptive capacity).

**3. What are the Impacts of India's Manufacturing Links in GVCs?**

- As for the impacts of India's GVC engagement, this research differentiates between economic and social upgrading. Here the economic upgrading is captured by three set of hypotheses considering the knowledge of expanding and strengthening GVC participation: (1) relationship between the growth rate of GVC participation and India's growth of domestic value-added embodied in exports (industry level) in the first part, (2) effect of GVC integration – as a buyer and a seller – on domestic value added (combining labour and capital with technology), also taking into account the mediating role of national policy (industry level) in the second part, and (3) influence of foreign investor characteristics and structural integration on productivity spillovers to domestic heterogeneous firms (considering firms' absorptive capacity and India's institutional variables) in the third part.

- In addition to economic upgrading, this research also looks at the impact of India's GVC engagement on social upgrading subject to the knowledge of turning GVC participation into sustainable development. This query will test the hypotheses about which GVC oriented industries have a higher demand for labour, such that integrating into GVCs in those sectors has a higher potential to create employment and increase household income (through labour value added). This analysis will provide an overview of measures that could be used to identify the impact on capital investment & stocks and employment by skill type. The measures would be categorised into two groups: indirect measures of social upgrading (the firm-level link between GVC participation and labour market outcomes), and direct measures of social upgrading (industry-level link from socio-economic accounts of world input-output tables).

Planning of this research should focus not only on the Indian economy as a whole, but also zoom into vital industries (based on the extensiveness of the GVC engagement), strategic segments therein, and individual value chains (as narrowly defined as the availability of quantitative and qualitative information allows).

## 5 RESEARCH METHODS

Here I have mentioned a summary of the probable methodologies (as well as data sources) available to carry out the GVCs' assessment and their content in this PhD study.

1. **Extent of India's Manufacturing GVC Links**

   - GVC Participation Using Gross Trade Data – comparing distributions of product-level exports with import values, volumes, and prices of the top traded products; informed classifications to extract information regarding meaningful clusters (value chain, technology); trade flows at the sub-national level to account India's degree of value transformation within the border.
   - Buyer's Perspective (Methods of Backward Links) – Share of intermediates in gross imports (range of imports, bundle of imported products, and countries), imported inputs embodied in exports (as percentage of gross imports and by source country), share of foreign value-added in gross exports (by source country), multinational's share of inputs from domestic suppliers in total inputs, domestic producer's share of imported inputs in total imported inputs, and Length of sourcing chains.
   - Seller's Perspective (Methods of Forward Links) – Share of intermediates in total exports (range of exports, bundle of exported products, and countries), domestic value added (% of gross value of output), domestic value added in gross exports of third countries, domestic value-added embodied in final foreign demand (% of GDP), domestic supplier's share of output to multinationals in total output, domestic supplier's share of exports in output, and Length of selling chains.

- GVC Participation (Methods from Macro to Micro) – GVC participation, position, and length indices, network metrics & visualisation, the role of services; an exchange between MNCs and domestic suppliers, domestic suppliers' & producers' shares in trade.

2. **Drivers of India's Manufacturing GVCs**

- Decomposition of Gross Export Growth – examining the level of significance of gross export growth (EXGR) onto its components (1) direct (intra-sector) domestic value-added embodied in gross exports (EXGR_DDC), (2) indirect (inter-sector) domestic value-added embodied in gross exports (EXGR_IDC), (3) re-imported domestic value-added (EXGR_RIM), and (4) foreign value-added embodied in gross exports (EXGR_FVA). The plausible estimation equation is:

$$\Delta. \ln EXGR_{st} = \alpha + \beta \Delta \ln EXGR\_DDC_{st} + \gamma \Delta \ln EXGR\_IDC_{st} + \delta \Delta \ln EXGR\_RIM_{st} + \theta \Delta \ln EXGR\_FVA_{st} + D_s + D_t + \varepsilon_{st}$$

Here $\Delta$ indicates first-order differences, while subscripts $s, t$ denote industrial sector and time. This regression control for sector $D_s$ and year $D_t$ fixed effects.

- Determinants of Sector GVC Participation – before analysing firm-level entry in GVCs, a preliminary evaluation could be made to estimate the impact of the policy determinants on sector GVC participation in India. The plausible estimation equation is:

$$\ln gvcpart_{st} = \alpha + \beta epz_{st} + \gamma open_{st} + \delta connect_{st} + \theta competitive_{st} + \tau invest_{st} + \varphi domest_{st} + D_s + D_t + \varepsilon_{st}$$

Here $gvcpart$ is the indicator of GVC participation; $epz$ is a dummy that equals 1 if India has an export-processing zone in an industrial sector, and 0 if not; $open$ denotes measures of openness and FDI; $connect$ denotes measures of connectivity to international markets (logistics, customs, and infrastructure); $competitive$ covers measures of competitiveness in unit labour costs and labour productivity; $invest$ captures drivers of investment (intellectual property protection, level of competition, administrative factors); and $domest$ denotes quality of domestic value chains and services infrastructure.

- Determinants of Firm-Level GVC Entry – following the literature on the firm-level determinants of importing to export, the model includes firm size, firm age, foreign ownership status, measures of workers' skills, productivity (depending on innovation capacity), and policy variables as determinants of GVC participation. The plausible estimation equation is:

$$gvc_{ist} = \alpha + \beta firm_{ist} + \gamma policy_{ist} + D_i + D_s + D_t + \varepsilon_{ist}$$

Here $gvc$ denotes a GVC indicator/dummy at the firm level, $firm$ is firm-level determinants of GVC integration, while subscript $i$ denotes firm, and $policy$ is policy determinants (institutional variables).

3. **Impacts of India's Manufacturing GVC Links**

- This objective can be explored through econometric analysis from several angles i.e. (a) whether the degree of structural integration in global value-added trade matters, (b) econometric analysis can be used to investigate how greater integration of India in GVCs as a buyer – as opposed to weaker integration as a seller (that is, more unbalanced GVC integration) – affects domestic value-added growth from gross exports, (c) the analysis can examine more closely the relationship between the growth of foreign value-added embodied in gross exports and the domestic value-added component, (d) it can look at the role of India's position in the value chain (upstreamness or distance to final demand), and (e) econometrics can be used to investigate the role of the domestic length of the sourcing chains. The model equation is:

$$\Delta \ln EXGR\_DVA_{st} = \alpha + \beta \Delta \ln backward_{st} + \gamma \Delta \ln forward_{st} + D_s + D_t + \varepsilon_{st}$$

Here $backward$ and $forward$ refer to different measures of backward and forward links of GVCs (according to hypotheses).

- GVC Links and Domestic Value Added – it focuses on the effect of GVC integration, as a buyer and a seller, on domestic value added, also taking into account the mediating role of national policy. Domestic value added is generated by combining labour with capital stock, and is dependent on a country's technology shifter. The technology shifter is assumed to be a function of international trade and innovation, which is consistent with the trade literature. The standard equation is:

$$\ln DVA_{st} = \alpha + \beta \ln GVC_{st} + \gamma \ln trade_{st} + \delta \ln capital_{st} + \theta emp_{st} + \tau \ln GVC_{st}.policy_{st} + D_s + D_t + \varepsilon_{st}$$

Here $DVA$ denotes domestic value added, $capital$ is capital stock, $emp$ is the nuber of employees, and $policy$ is policy variables. $GVC$ captures the different structural measures of GVC participation, which enter the function as part of technology shifter. In addition to GVC participation, $trade$ is a measure of final goods trade to separate the potential positive GVC effect from the simple positive effect of trade openness. The model also includes different fixed effects. The last are included to capture innovation that is part of the technology shifter.

- GVC Participation and Firm-Level Productivity – The method focuses on the within industry impact of foreign output share on domestic firm productivity and the role of mediating factors; Similarly, the analysis can be used to examine the effect of GVC participation of an industry on a firm's productivity. The baseline equation is:

$$\ln LP_{ist} = \alpha + \beta GVC_{ist} + \gamma GVC_{ist}.MF_{ist} + \delta \ln capint_{st} + D_i + D_s + D_t + \varepsilon_{ist}$$

Here $LP$ denotes labour productivity for domestic firm in sector $s$ at time $t$; $capint$ denotes capital intensity. $MF$ is the set of mediating factors (measures of spillover potential by the foreign firm, measures of absorptive capacity in India, and measures of national characteristics and institutions).

- Indirect Measures of Social Upgrading – descriptive statistics of manufacturing averages of the number of employees, wages and salaries, wage rate (wages and salaries divided by the number of employees), or labour share (wages and salaries as a percentage of value added). Besides this, the labour market indicators are regressed on indicators of GVC involvement while controlling for other factors, such as region and gross domestic product. I can also run pooled regressions controlling for industry fixed effects to see which industries have more labour-market-enhancing outcomes conditional on GVC involvement.

- Direct Measures of Social Upgrading – Labour Content of Gross Exports, Labour Component of Domestic Value Added in Exports, Jobs Sustained by Foreign Final Demand, Jobs Generated by Foreign Trade in GVCs

**Data Sources:** Production data – Industry-level and firm-level (Enterprise surveys or other firm-level surveys); Gross trade data (Comtrade, WITS), categorized using informed classifications (broad economic category, parts and components, technical classifications); International I-O data (WIOD, TiVA, World Bank Export of Value Added database) and National I-O data. World I-O table (Timmer et al. 2014) is based on national supply- & use-tables A, combined with time-series on v and F from national accounts statistics and bilateral trade data from official statistical sources (by use category). Socio-economic accounts include data on hours worked and wages by three skill types (educational attainment levels) and capital.

![World input-output table to GVC cost-share table conversion, with formula $G = v(I-A)^{-1}F$ — Leontief's trick: compute value added in all industries associated to final demand for a specific product]

## 6 TENTATIVE CHAPTERS

By literature review, research is required to obtain a more precise & an exclusive understanding of India's circumstances under which more inclusive and sustained economic as well as social upgrading in GVCs can be attained for all industry and workers. As part of this study, the chapters in this PhD dissertation would contribute some research findings that shed on several key challenges that can be addressed with the tools described in the previous sections include the following:

- Chapter 1: Multidimensional Assessment of India's GVC Linkages

    1. GVC Participation Using Gross Trade Data
    2. Buyer's & Seller's Perspectives: Backward & Forward Links
    3. GVC Participation: Methods from Macro to Micro

- Chapter 2: Drivers of India's GVC Links

    1. Determinants of industry-level GVC participation
    2. Determinants of firm-level entry in GVCs

- Chapter 3: Impacts of GVCs on India's Economic and Social Prosperity

    1. Economic Upgrading of GVC Links
    2. Social Upgrading of GVC Links